\documentclass[%
 reprint, 
 superscriptaddress,
 amsmath,amssymb,
 aps,
]{revtex4-2}

\usepackage{graphicx}
\usepackage{dcolumn}
\usepackage{bm}
\usepackage{siunitx}
\usepackage{gensymb}

\begin{document}

\preprint{APS/123-QED}

\title{Inducing Skyrmion Flop Transitions in Co$_8$Zn$_8$Mn$_4$ at Room Temperature}



\author{Simon A. Meynell}

\altaffiliation{These authors contributed equally.}
\affiliation{Department of Physics, University of California, Santa Barbara 93106 CA, USA}%

\author{Yolita M. Eggeler}%
\altaffiliation{These authors contributed equally.}

\affiliation{Materials Department and Materials Research Laboratory, University of California, Santa Barbara CA 93106 USA}%
\affiliation{Laboratory for Electron Microscopy, Karlsruhe Institute of Technology, D-76131 Karlsruhe, Germany}%

\author{Joshua D. Bocarsly}%
\affiliation{Materials Department and Materials Research Laboratory, University of California, Santa Barbara CA 93106 USA}%

\author{Daniil A. Kitchaev}
\affiliation{Materials Department and Materials Research Laboratory, University of California, Santa Barbara CA 93106 USA}%

\author{Bailey E. Rhodes}
\affiliation{Materials Department and Materials Research Laboratory, University of California, Santa Barbara CA 93106 USA}%

\author{Tresa M. Pollock}
\affiliation{Materials Department and Materials Research Laboratory, University of California, Santa Barbara CA 93106 USA}%

\author{Stephen D. Wilson}
\affiliation{Materials Department and Materials Research Laboratory, University of California, Santa Barbara CA 93106 USA}%

\author{Anton Van der Ven}
\affiliation{Materials Department and Materials Research Laboratory, University of California, Santa Barbara CA 93106 USA}%

\author{Ram Seshadri}
\affiliation{Materials Department and Materials Research Laboratory, University of California, Santa Barbara CA 93106 USA}%
\affiliation{Department of Chemistry and Biochemistry, University of California, Santa Barbara CA -93106 USA}

\author{Marc De Graef}
\affiliation{Department of Materials Science and Engineering, Carnegie Mellon University, Pittsburgh, PA 15213, USA}

\author{Ania Bleszynski Jayich}
\email{ania@physics.ucsb.edu}
\affiliation{Department of Physics, University of California, Santa Barbara 93106 CA, USA}

\author{Daniel S. Gianola}
\email{gianola@ucsb.edu}
\affiliation{Materials Department and Materials Research Laboratory, University of California, Santa Barbara CA 93106 USA}%

\date{\today}

\begin{abstract}

Magnetic skyrmions are topologically-protected spin textures that manifest in certain non-centrosymmetric ferromagnets under the right conditions of temperature and field. In thin film skyrmion hosts, demagnetization effects combined with geometric confinement can result in two distinct types of spin textures: those with their axis of symmetry in the plane of the film (IP) and those with their axis pointing out-of-plane (OOP). Here we present Lorentz transmission electron microscopy evidence in conjunction with numerical modeling showing a flop transition between IP and OOP skyrmions in Co$_8$Zn$_8$Mn$_4$ at room temperature. We show that this skyrmion flop transition is controllable \textit{via} the angle of the external field relative to the film normal and we illustrate how this transition depends on thickness. Finally, we propose a
skyrmion-writing device that utilizes the details of this transition.
\end{abstract}

\maketitle

\section{Introduction}

Since their postulation in 1994 \cite{BOGDANOV1994_thermodynamicallystablevortex}, the vortex-like quasiparticles known as magnetic skyrmions, have been the focus of efforts to create high-information-density, nanoscale digital storage devices \cite{Parkin2008_racetrackreview}. 
The skyrmion is a nanoscale \cite{Wang2018_skyrmionsize} cylindrical magnetic field twist whose spins wind in a topologically stable vortex structure \cite{Nagaosa2013_TopologicalProperties}. 
The ability to manipulate skyrmions even at very low current density \cite{Purnama2015_guidedcurrentskyrmion,yu2020motion} has made them attractive for digital memory storage applications due to the consequent low-power requirements \cite{Fert2013_skyrmionontrack,yu2012skyrmion,peng2021dynamic}.
Furthermore, topological protection can result in an extremely long lifetime, which is one of the prerequisites for a bit in a digital memory element \cite{Bessarab2018_SkyrmionLifetime}.
Another essential requirement for a useful digital memory element is the ability to controllably write and delete bits on demand.
The challenge of writing and deleting skyrmions is a major obstacle to developing skyrmion-based devices and is an ongoing scientific and technological area of investigation \cite{Romming2013_WritingAndDeleting,Wang2022_ElectricalManipulationSkyrmions,QIU2022_writingskyrmionSpecificPosition}.

Many of the proposed methods for writing, or nucleating, skyrmions involve large current pulses that ease the local nucleation of a skyrmion \cite{Gan_2018_Skyrmion_Injection,Quessab2022_ZeroFieldNucleation,Wang2022_ElectricalManipulationSkyrmions,Jiang2015_BlowingSkyrmionBubbles}.
Despite many successful demonstrations of skyrmion injection by these methods, developing defect-free nucleation mechanisms that do not rely on driving large current pulses through the microstructure would broaden the applicability of skyrmion-based devices and enable new device architectures.

Central to the physics of skyrmion nucleation and their stability is a chiral term arising in the spin-spin Hamiltonian known as the Dzyaloshinskii-Moriya interaction (DMI) that arises in the presence of broken inversion symmetry.
This term can stabilize a wide variety of long-range chiral magnetic orders such as magnetic helices, conical structures,  and skyrmions \cite{Dzyaloshinksii58,Moriya60,BOGDANOV1994_thermodynamicallystablevortex,everschor-sitte_review2018,zhou_review2018}. 
Among skyrmions in thin films stabilized by bulk DMI, there exist two primary structure types \cite{Leonov2020_stabilityIPvsOOP} — skyrmions with their axis of symmetry within the thin-film plane \cite{Moon2019_in-planeskyrmiontheory,Meynell2017_MnSiInPlane,Yokouchi2015_InPlaneMnSiSkyrmion}, which we refer to as in-plane (IP), and the more conventional type whose axis of symmetry lies parallel to the film normal, which we refer to as out-of-plane (OOP) \cite{karube2016robust}. 
While there is no fundamental difference between these skyrmion types in bulk materials, the large shape anisotropy of thin films causes these two skyrmion types to differ in their energy densities, regions of phase stability, and behaviors with respect to external stimuli.

\begin{figure*}
\includegraphics[scale = 0.65]{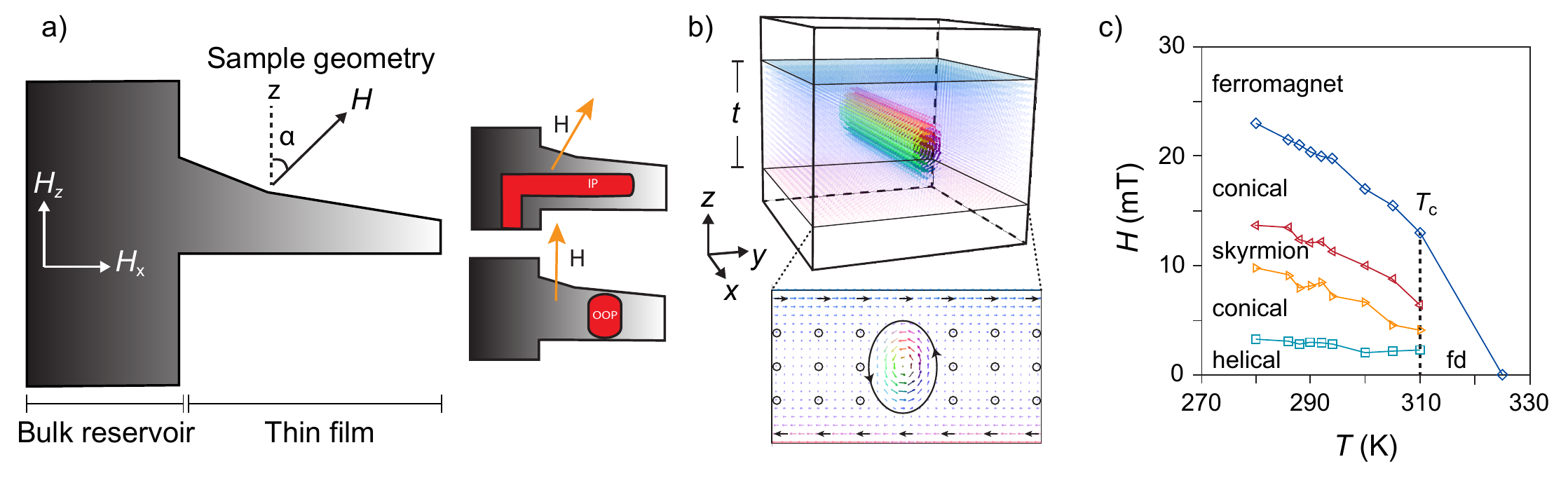}
\caption{\label{fig:1introduction}
(a) Schematic of the experimental geometry. The angle, $\alpha$, is defined as the angle between the film normal and the applied field. 
(b) Schematic of an in-plane skyrmion in a semi-infinite film of thickness $t$. The z direction of the coordinate system is normal to the film; x and y lie in the film plane. 
(c) Magnetic phase diagram for bulk Co$_8$Zn$_8$Mn$_4$ as a function of applied magnetic field, $H$, and temperature, $T$. The helical, conical, and skyrmion magnetic orderings dominate at low temperature but terminate in the fluctuation-disordered (fd) regime as temperature increases. At high field and high temperatures, the material saturates into a ferromagnetic ordering. }
\end{figure*}

The relative stabilities of OOP and IP skyrmions, as well as the transitions between the two types, have been minimally explored.
The field direction plays a large role in determining the stable orientation of the skyrmion; a large field tends to destabilize skyrmions with their axis of symmetry orthogonal to the field direction.
For example, in [Pt/Co/Ta] multilayers, it was demonstrated that tilting the applied field away from the thin film normal destabilizes out-of-plane skyrmions and preferentially stabilizes a helical state  \cite{Zhang2018_NeelSkyrmionIPField}. 
Separately, it has been shown that in non-centrosymmetric magnets, the field angle controls a transition between a helical state and OOP skyrmions \cite{Wang2021_stimulated_nucleation_centrosymmetric}.


We posit that in addition to a transition between helical and OOP skyrmions, the field angle can be used to control the skyrmion axis of symmetry.
By sweeping through the field angle, skyrmions may be reoriented between IP and OOP via a flop-transition, similar to the one predicted in Ref. \cite{Vlasov_2020_skyrmionFlop}.
Most proposed skyrmion nucleation methods to date involve the injection of OOP skyrmions because they have the preferred small size and well-established methods of current manipulation necessary for racetrack memory storage. 
However, if skyrmion flop transitions could be controllably induced, then devices that nucleate IP skyrmions could be used as a source for OOP skyrmions. 


In this contribution, we report direct observations of flop transitions between in-plane and out-of-plane skyrmions by systematically varying the magnetic field \emph{in situ} within a transmission electron microscope (TEM).  
Our specimen geometry enables us first to nucleate in-plane skyrmions from a bulk topological charge reservoir and then exploit a flop transition to switch between in-plane and out-of-plane skyrmions controllably.
We use a combination of a tilted magnetic field and a variable film thickness to elucidate the stability limits of the two skyrmion geometries and, consequently, identify the conditions for the flop transition.  
Informed by these observations and corroborating magnetic simulations, we propose a novel and potentially scalable racetrack memory component that benefits from both orientation types of skyrmions.

\section{Methods}

\par The bulk skyrmion host Co$_x$Zn$_y$Mn$_z$ ($x + y + z = 20$), crystallizing in the $\beta$-Mn-type structure, generally supports skyrmions across a wide range of stoichiometries. 
Previous investigations have revealed that room temperature skyrmion formation is limited to a narrow composition window centered around Co$_8$Zn$_8$Mn$_4$ \cite{Tokunaga2015_A_New_Class,Bocarsly19_CoZnMn,karube2016robust}. 
A polycrystalline ingot of this composition was prepared from elemental starting materials of Co powder, Zn shot, and Mn pieces, which were weighed with an 8:8:4 stoichiometry totaling a mass of 5 g following a procedure reported in previous investigations \cite{Tokunaga2015_A_New_Class,Bocarsly19_CoZnMn}. 
The metals were sealed in an evacuated silica ampule and heated to 1,000$^\circ$C for 25 hr and then cooled to 925$^\circ$C at a rate of 1$^\circ$C$\cdot$hr$^{-1}$. 
Finally, the sample was homogenized at 925$^\circ$C for 48 hr. before quenching to room temperature in a water bath. 
The magnetic phase diagram near room temperature of the bulk sample was inferred using AC magnetic susceptibility measurements, as has previously been performed in $\beta$-Mn structure skyrmion hosts \cite{Tokunaga2015_A_New_Class, Kautzsch2020}. 
AC susceptibility as a function of applied DC magnetic field (0\,$< H <$\,50\,mT) was measured using a Quantum Design MPMS3 Magnetometer at fixed temperatures ranging from 310\,K to 274\,K. 
The phase boundaries between helical, conical, skyrmion, conical, and ferromagnetic states were determined from alternating local maxima and minima in plots of $d \chi$/$d H$ \emph{vs.} $H$ as previously described \cite{Kautzsch2020}. 

The Co$_8$Zn$_8$Mn$_4$ TEM specimen was prepared by micromachining from the polycrystalline ingot, using a focused ion beam incorporated within an FEI Helios Dual-beam Nanolab 600. 
A 15 $\mu$m x 7 $\mu$m rectangular specimen was extracted from the Co$_8$Zn$_8$Mn$_4$ bulk sample with an initial thickness of 400-500 nm and attached to a TEM Cu support grid for further thinning down to electron transparency at an average thickness of \SI{90}{\nano\meter}. 
The thinning procedure resulted in a wedge-shaped specimen geometry attached to a bulk reservoir as schematically demonstrated in (Fig. \ref{fig:1introduction}(a)), with the thickness of the thin film ranging from \SI{160}{\nano\meter} to \SI{90}{\nano\meter}, measured from cross-sectional imaging using secondary electrons at 10 keV.  


Lorentz TEM (LTEM) was used to observe the magnetic structures as a function of the magnetic field using two approaches. 
Fig. \ref{fig:1introduction}(a) shows the sample and field geometry for this experiment. 
The first experiment used a fixed specimen tilt angle $\alpha$ and varied the applied magnetic field strength, $H$, by changing the objective lens strength on a FEI ThemIS 60-300 TEM operating in Lorentz mode at 200 kV. 
The magnetic field as a function of the objective lens excitation was calibrated using a Hall probe TEM holder. 
The second set of experiments varies $\alpha$ at a fixed $H=\SI{120}{\milli\tesla}$ generated by a weakly excited objective lens operating at low magnification in a Thermo Fisher Talos 200FX TEM operating at 200 keV.
The behavior of the magnetic structures was observed by obtaining a series of Fresnel images at large defocus conditions in the range of several mm. 
A through-focus image series was reconstructed to generate magnetization maps using an electron wave phase retrieval method that employs the transport-of-intensity equation \cite{volkov2002new}.  
This approach uses the defocused image series at under- and overfocused conditions (centered at zero defocus) to recover a phase map, which in turns enables the calculation of phase gradients, which are related to the in-plane components of the integrated magnetic induction \cite{aharonov1959significance}.

We use MuMax3 \cite{mumax3_citation} to model the impact of thin-film anisotropy on the energetics of IP and OOP skyrmion states. 
The material parameters used in the calculations are $D_{\textrm{bulk}} = \SI{5.3e-4}{J/\meter^2}$, $M_{\textrm{sat}} = \SI{410}{kA/\meter}$, and $A_{\textrm{exch}} = \SI{4.6e-12}{J/\meter}$. 
The size of the numerical system is $32\times32\times32$ cells with a cell size of $\SI{7}{\nano\meter}$. 
Periodic boundary conditions are enforced in the XY-plane.

To calculate the competing energetics of OOP and IP skyrmions, we initialize their magnetization using an idealized Bloch skyrmion, where the magnetization points along z (x) for OOP (IP) skyrmions. 
Figure  \ref{fig:1introduction}(b) shows an example of a simulated IP skyrmion. 
For OOP skyrmions, $H$ is set along the z-direction, and for IP skyrmions, $H$ is tilted at an angle $\alpha = 15 \degree$ from the z-axis towards the x-axis. 
The magnetic structure is then relaxed using the conjugate gradient method to arrive at a stable configuration of spins. 
Then $\alpha$ is gradually varied in steps of $0.25\degree$, with a conjugate gradient minimization of the spin structure under each condition.
We repeat this process for different field magnitudes ($\SI{50}{\milli\tesla} - \SI{150}{\milli\tesla})$ and film thicknesses ($\SI{60}{\nano\meter} - \SI{200}{\nano\meter}$).

\section{Results and Discussion}

As has previously been observed with bulk magnetic measurements and LTEM measurements, Co$_8$Zn$_8$Mn$_4$ can host skyrmions at room temperature under a moderate applied magnetic field. 
Figure \ref{fig:1introduction}(c) depicts the bulk magnetic phase diagram of our sample of Co$_8$Zn$_8$Mn$_4$ inferred using AC magnetic susceptibility, which is consistent with previous reports \cite{Tokunaga2015_A_New_Class}. 
The skyrmion magnetic phase competes with other chiral magnetic orderings: at very low applied fields, a helical magnetic phase is observed. 
The helical phase transitions to a conical phase upon application of larger fields in which the helical moments cant in the direction of the applied field before eventually reaching a fully field-polarized ferromagnetic state. 
Skyrmions exist as the stable phase in a small pocket at moderate magnetic fields ($\approx $\SI{10}{\milli\tesla}) within the conical phase space. 
Here we directly confirm the existence of skyrmions in thin-films of Co$_8$Zn$_8$Mn$_4$ via LTEM imaging.

To describe the geometry of skyrmions within a thin film of thickness $t$ in Co$_8$Zn$_8$Mn$_4$, we adopt the coordinate system shown in Figure \ref{fig:1introduction}(b), where the $z$-axis is parallel to the film normal and the $x$-axis points along the in-plane component of the external field. 
For example, an in-plane skyrmion (IP) has its axis within the $x-y$ plane of the film as depicted in Fig. \ref{fig:1introduction}(b). 
In the bulk limit, $t \rightarrow \infty$, the skyrmion axis of symmetry is parallel to $H$.
We note that magnetocrystalline anisotropy is not considered because of the polycrystalline nature of our films: in our images, we see multiple grain boundaries, and the magnetic textures are uninterrupted across them, indicating shape anisotropy dominates.

In a thin sample, the shape anisotropy leads to a large demagnetizing field that changes the relative stability of IP and OOP skyrmions as a function of $\alpha=\arctan(H_z/H_x)$. 
Generally, the demagnetizing field in a thin film suppresses the out-of-plane component of the magnetization \cite{dove1967demagnetizing}. 
Since skyrmions have a net magnetization along their high-symmetry axis, this phenomenon is expected to decrease the critical angle $\alpha^*$ of the skyrmion flop transition.
This trend persists as long as the film thickness $t$ is larger than the skyrmion diameter (given by the material's helical pitch $\lambda$). 
For $t < \lambda$, the IP phase is suppressed by confinement. 
We leverage this confinement-induced shape anisotropy effect to study the skyrmion flop transition using the variable-thickness sample geometry illustrated in Figure \ref{fig:1introduction}(a). 
By varying $\alpha$ and measuring over different regions of the sample, we can explore the effects of both thickness and angular dependence for the skyrmions observed in the Co-Zn-Mn sample.


\subsection{Constant angle skyrmion nucleation}
For novel data storage technologies, the OOP skyrmion has been the preferred focus of investigations because of its small diameter, allowing for high information density, and the existence of established techniques for current-driven spatial manipulation.
Most observations of skyrmion nucleation are of the OOP variety.
Here, we inject IP skyrmions from a bulk reservoir into a thin film and subsequently nucleate OOP skyrmions using the OOP-IP flop transition.
This technique forms the basis for our proposed hybrid device discussed in section \ref{sec:Device}.


\begin{figure}
\includegraphics[scale = 0.75]{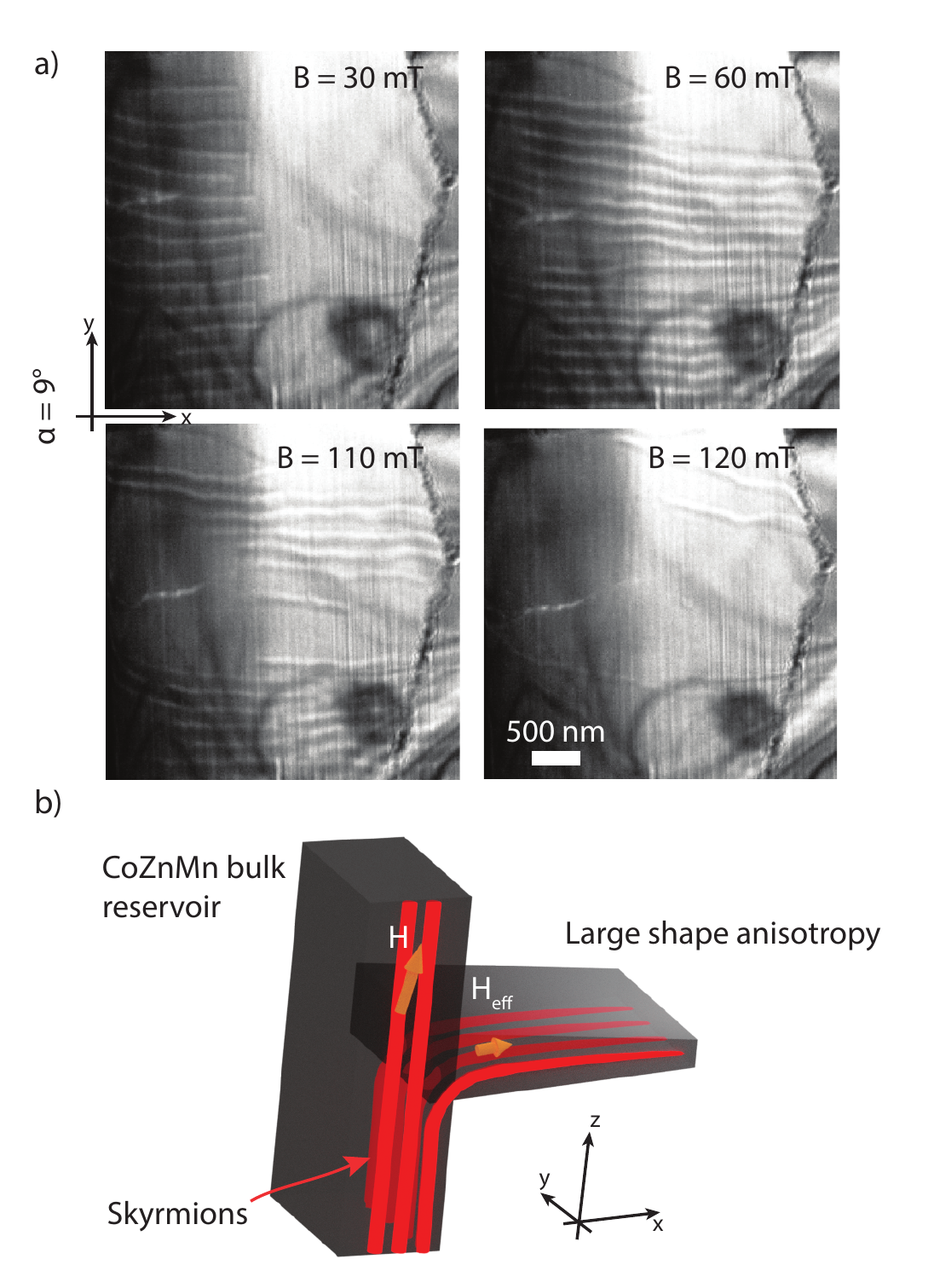}
\caption{\label{fig:IP_nucleation} (a) LTEM sequence of images of a wedge of CoZnMn with the applied field held at a constant angle ($\alpha = 9\degree$) while increasing magnetic field strength B. For small fields, the in-plane skyrmions only extend part way into the film. As the field strength increases, the skyrmions contract in size and thus can reach into the thinner regions of the film. (b) A proposed mechanism for in-plane skyrmion nucleation. The large shape anisotropy in the wedge section suppresses the out-of-plane component of the tilted external field, leaving only the in-plane component, stabilizing the in-plane skyrmions.
}
\end{figure}


Figure \ref{fig:IP_nucleation} shows a series of defocused LTEM images of the wedge sample at varying field strengths with a fixed tilt angle $\alpha = 9\degree$. 
As will be shown later, this $\alpha$ is large enough such that IP skyrmions will always be favored in the geometry of our sample. 
The left side of the images is near to the bulk material with the foil thinning towards the right. 
At a low field of $H = \SI{30}{\milli\tesla}$, we observe stripes with a mean spacing of $\sim\SI{140}{\nano\meter}$ emerging from the bulk reservoir on the left and extending toward thinner regions of the film as the field increases.
The stripes have two possible interpretations: this texture is either a 1D lattice of IP skyrmions, or it is a helical/conical state with $Q\cdot B_x = 0$. 
We infer that these are IP skyrmions because the in-plane component of the film is pointing along the skyrmion axis. 
Were this a Bloch-type helical/conical phase, it would be energetically favorable for the wave vector of the helix to point along the in-plane field component (\emph{i.e.} the stripes would appear perpendicular to the direction in which we observe them). 
We further compare the reconstructed magnetization to our data and find good agreement between a simulated in-plane skyrmion and the LTEM image.

For low fields, IP skyrmions extend through the thicker regions of the film but stop when the film thickness becomes small relative to the skyrmion size as shown in the $H = \SI{30}{\milli\tesla}$ subimage of Fig. \ref{fig:IP_nucleation}(a). 
As we increase the field to $\SI{60}{\milli\tesla}$ we see an increase in skyrmion density, which we attribute to the effect of decreasing skyrmion size with increasing external field \cite{wang2018theory}. 
The reduction of skyrmion size also allows for them to extend further into thinner regions of the film. 
At fields of $H = \SI{110}{\milli\tesla}$ and $\SI{120}{\milli\tesla}$ (which correspond to $H_z$\,=19\,mT), IP skyrmions begin to disappear altogether, saturating into either a ferromagnetic or an out-of-plane cone state, as is suggested by the magnetic phase diagram in Figure \ref{fig:1introduction}(c).

The proposed mechanism for the in-plane skyrmion nucleation we observe is shown in Fig. \ref{fig:IP_nucleation}(b). 
Here, the skyrmions that run vertically through the bulk reservoir are interrupted, where they transition from the bulk and are drawn into the wedge. 
Based on this model, we propose that the reservoir acts as a topological charge bath for the attached thin-film wedge, providing a means to overcome the large energy barrier associated with skyrmion nucleation. 
Having established the nucleation and migration of in-plane skyrmions with increasing field magnitude but fixed orientation, we next investigate the behavior of the skyrmion flop transition as a function of film thickness and external field angle.

\begin{figure}
\includegraphics[scale = 0.6]{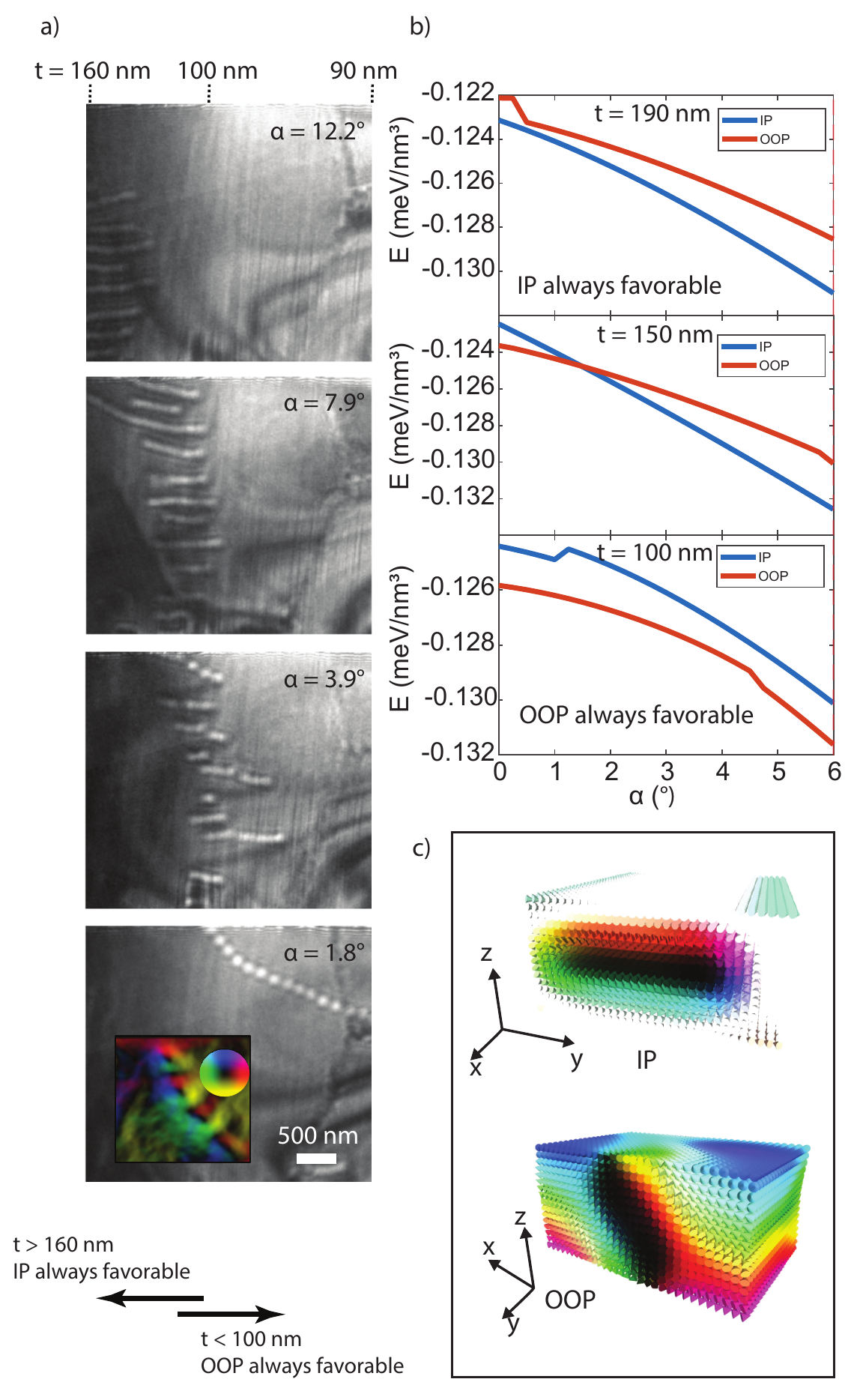}
\caption{\label{fig:skyrmion_flop} 
(a) LTEM image sequences of a CoZnMn TEM wedge sample for an out of plane field of $B = \SI{120}{\milli\tesla}$ and varying film angle, $\alpha$.
At high angles, the perpendicular component of the field suppresses the OOP phase and the IP is present in the thicker regions of the film as indicated by the lack of OOP on the left side of the film and the lack of IP on the right side of the film for all values of $\alpha$. 
Close to $\alpha = 4\degree$, the IP undergoes a flop-transition into an OOP phase that is only stable in the thinner regions of the wedge. 
The colormap in the $\alpha = 1.8\degree$ image shows the phase reconstruction for OOP skyrmions as detailed in the supplemental information. The color indicates the direction of the IP magnetization and confirms the Bloch character of the skyrmions.
(b) Simulations performed via mumax3 showing the energy dependence as a function of $\alpha$ for different thicknesses of the wedge. 
For intermediate thicknesses ($t = \SI{150}{\nano\meter}$), the simulations reveal a crossover angle where the IP phase will flop to the OOP variety, which agrees with our experimental observation that the skyrmions flop in the intermediate film thickness region.
(c) Equilibrium simulations of IP and OOP skyrmions with a tilted field in a geometrically confined thin film.
}
\end{figure}

\subsection{Skyrmion flop transition}

The principle of skyrmion axis reorientation was predicted by Vlasov et al. in Ref. \cite{Vlasov_2020_skyrmionFlop}, where the authors showed that skyrmions can change orientation via an intermediate state with two mutually orthogonal skyrmions. 
Figure \ref{fig:skyrmion_flop}(a) shows a series of LTEM images that demonstrates the flop transition as a function of the tilt angle and film thickness.
The thickness of the TEM foil was measured from cross-section orientation images acquired by SEM and is labeled by the vertical dotted lines at the top of Figure \ref{fig:skyrmion_flop}(a).  
The leftmost part of the image is the thickest region of the foil, which connects to the bulk reservoir.

An applied field of $\SI{120}{\milli\tesla}$ is aligned along the film normal and swept between angles of $\alpha= 15\degree$ - $0\degree$. 
A video demonstrating the magnetic texture evolution as the field is continuously tilted is provided in the online supplementary material.
For high tilt angle, $\alpha = 12.2\degree$, IP skyrmions nucleate at the thicker side of the film, close to the bulk reservoir (Fig. \ref{fig:skyrmion_flop}(a)). 
In these thicker regions of the film, all skyrmions observed are in-plane. 
The IP skyrmions exhibit capped ends, similar to those observed by Refs \cite{zheng2018experimental,Birch2022_skyrmionBlochPoints}. 
For $\alpha = 7.9\degree$, the IP skyrmions move into thinner regions of the film.
As the field angle approaches that of the film normal at $\alpha = 3.9\degree$, we approach the critical angle, $\alpha^*$.
We find that as $\alpha$ decreases, the IP skyrmions shift toward thinner regions of the film, where the transition between the two types is observed as a mixture of IP and OOP skyrmions in the image. 
For the small angle of $\alpha = 1.8\degree < \alpha^*$, the image shows only small, pearl necklace-type OOP skyrmions. 
 
The magnetization map inset in the $\alpha = 1.8\degree$ image shows a vortex of magnetic moments, confirming the chirality and Bloch-like character of the OOP skyrmions. 
The magnetization maps are generated from the LTEM images via a procedure detailed in the SI.

Micromagnetics calculations shown in Figure \ref{fig:skyrmion_flop}(b) plot the expected impact of sample thickness on the skyrmion orientation by calculating the energy density of the two orientations of skyrmions as a function of applied field angle for three different film thicknesses—$190$, $150$ and $\SI{100}{\nano\meter}$. 
The results of the numerics agree with the experimental data in Fig. \ref{fig:skyrmion_flop}(a): the calculations predict IP to be the stable phase for large film thicknesses while for thin thicknesses OOP becomes the favorable phase due to confinement effects.
Notably, for intermediate thicknesses, the micromagnetics predicts a critical angle, $\alpha^*$, where the energies of the two orientations are the same, and the skyrmion flop transition can therefore occur.
Examples of the equilibrium magnetization distributions used to generate the phase diagram and curves are depicted in \ref{fig:skyrmion_flop}c, for both OOP and IP skyrmions, respectively.

Interestingly, all OOP phases observed in these experiments were arranged together into a 1D chain, suggesting the presence of an attractive skyrmion-skyrmion interaction.
One possible explanation for this morphology could be stabilization via the `zip-locking' mechanism proposed by Vlasov \emph{et al.} \cite{Vlasov_2020_skyrmionFlop}, where an IP guides the OOP skyrmions along an axis.

\begin{figure}
\includegraphics[scale = 0.75]{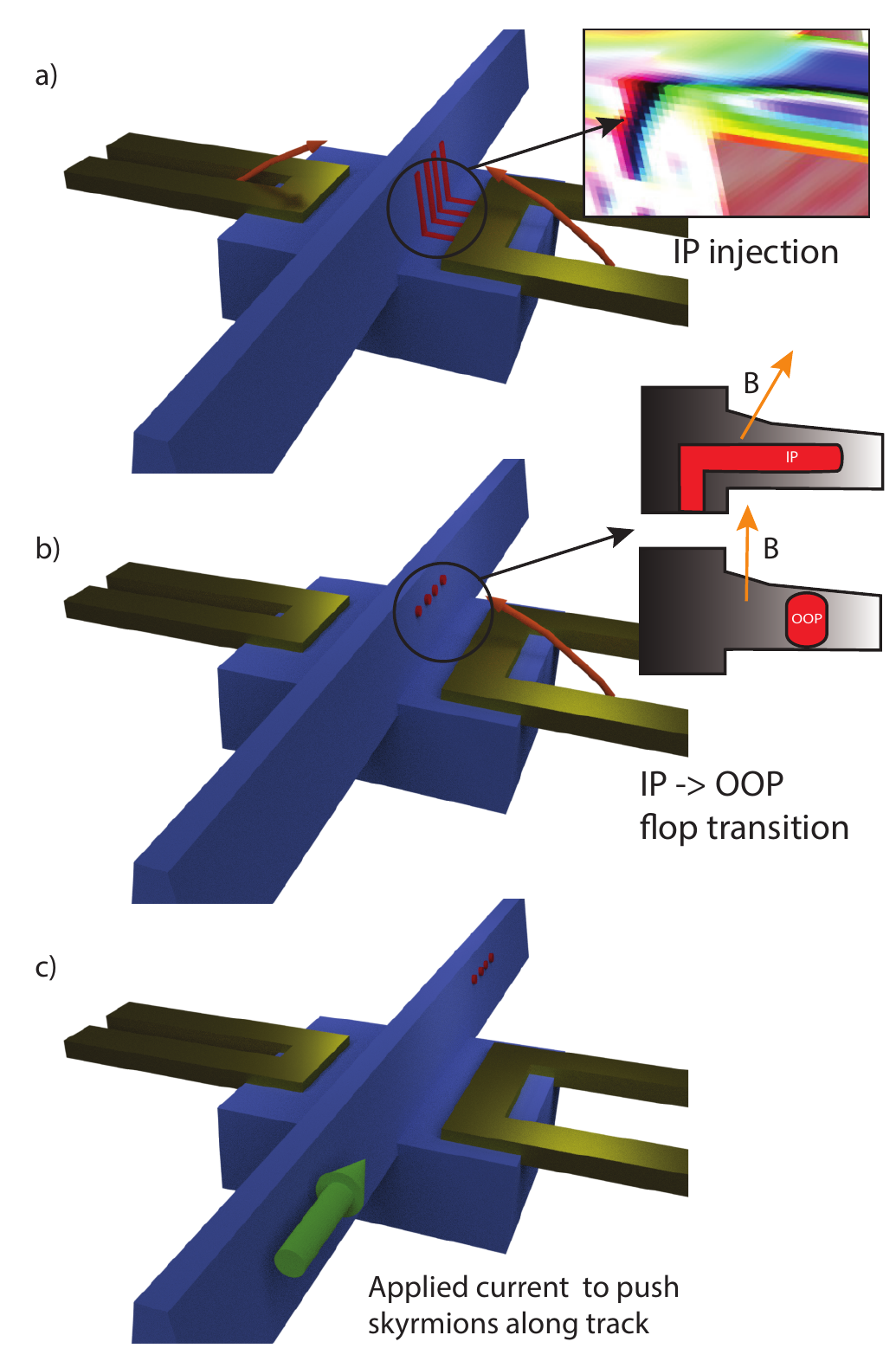}
\caption{\label{fig:device} A schematic for a skyrmion writing component utilizing the mechanisms demonstrated in this work. 
The blue material represents the Co$_8$Zn$_8$Mn$_4$ material, the gold represents a current carrying stripline, the orange arrows represent an applied field generated by the striplines and the red represents a skyrmion core.
(a) The first step is to inject in-plane skyrmions through the mechanism described in Fig. \ref{fig:IP_nucleation}(a). Gold wires may be used to trim the field strength and angle. 
(b) The injected IP skyrmions may be transformed into OOP skyrmions via the flop transition shown in Fig. \ref{fig:skyrmion_flop}(a).
The inset shows how the field is tilted with respect to the wedge to induce the IP-OOP flop.
(c) After the OOP skyrmions are formed, they can be manipulated along the track through an applied current.
}
\end{figure}

Wang et al. \cite{Wang2021_stimulated_nucleation_centrosymmetric} demonstrate a skyrmion nucleation method similar to the one illustrated here with a few key differences.
First, they are studying a centrosymmetric material without DMI, and second, they show a transition between magnetic helix to skyrmion-in-helix states.

\subsection{Device governed by skyrmion flop transition} \label{sec:Device}

We have shown two properties of the Co$_8$Zn$_8$Mn$_4$ alloy shaped as a thin wedge sample in a tilted field. 
First, we have observed the generation of IP skyrmions near the reservoir-film interface (Figure \ref{fig:IP_nucleation}). 
Second, we have demonstrated that depending on the host film thickness, the IP skyrmions can undergo a flop transition into the OOP orientation induced by sweeping the field angle (Figure \ref{fig:skyrmion_flop}).
Based on these two principles, we propose a new concept for a skyrmion generation writing component in a racetrack memory device shown in Figure \ref{fig:device}.
The operating principle of the device relies on the skyrmion injection method from a bulk reservoir, followed by a field-tilt-induced flop transition from IP to OOP in a thin region of the attached wedge-shaped wire. 
To control the field magnitude and the field angle, two gold striplines close to the region of interest are proposed.

The first step in generating skyrmions would be to apply a tilted field far from the critical angle for the flop transition. 
Using 30 mT as in Figure \ref{fig:IP_nucleation}(a) will be sufficient to produce a semi-localized IP phase in the wedged portion of the device. 
Subsequent trimming of the field may change the field angle or magnitude, flopping the IP into OOP, which is the desired skyrmion configuration for high densities of nanobits, schematically shown in Figure \ref{fig:skyrmion_flop}(b).

Finally, with the stable OOP skyrmions generated along the device, a current may be applied along the wedge illustrated by the green arrow in Figure \ref{fig:device}(c). 
Due to skyrmion motion under an applied current \cite{Masell2020_STTdrivenmotion,Hrabec2017_CurrentInducedSkyrmionGeneration,Wang2022_ElectricalManipulationSkyrmions}, the nucleated skyrmions can be transported along the track, for example to a reader head.
This reservoir-wedge configuration then acts as a `skyrmion-writer' element of a larger skyrmion memory device.

\section{Conclusions}

We demonstrate the existence and regions of stability for in-plane skyrmions, IP, in thin-film Co$_8$Zn$_8$Mn$_4$ as well as show the field-angle induced skyrmion flop transition from an IP to an OOP spin-texture. 
In-plane skyrmions, while less studied than their out-of-plane counterparts, provide a path toward controlled nucleation of skyrmions in a geometrically confined reservoir-wedge configuration and a room-temperature device. 
We show that the flop transition is repeatable across two different devices and that control over field angle and wire geometry provides a method of switching between the types.
Lastly, we propose a component of a larger race-track device utilizing these principles that could pave the way for current-controlled writing of skyrmion bits in drastically reduced scales and low-power data storage components.

\section{Acknowledgements}

This work was supported by the Materials Research Science and Engineering Center (MRSEC) program of the National Science Foundation (NSF) through DMR-1720256 (IRG-1), and employed the shared facilities of the MRSEC at UC Santa Barbara, a member of the Materials Research Facilities Network. Work at the Molecular Foundry was supported by the Office of Science, Office of Basic Energy Sciences, of the U.S. Department of Energy under Contract No. DE-AC02-05CH11231. SAM acknowledges the support of the Natural Sciences and Engineering Research Council of Canada (NSERC), [AID 516704-2018] and the UCSB Quantum Foundry.

\bibliography{skyrmion_flop.bib}

\end{document}


\title{Supplemental Information.}
\maketitle

\section{LTEM Reconstruction}

\begin{figure}

\includegraphics[scale = 0.75]{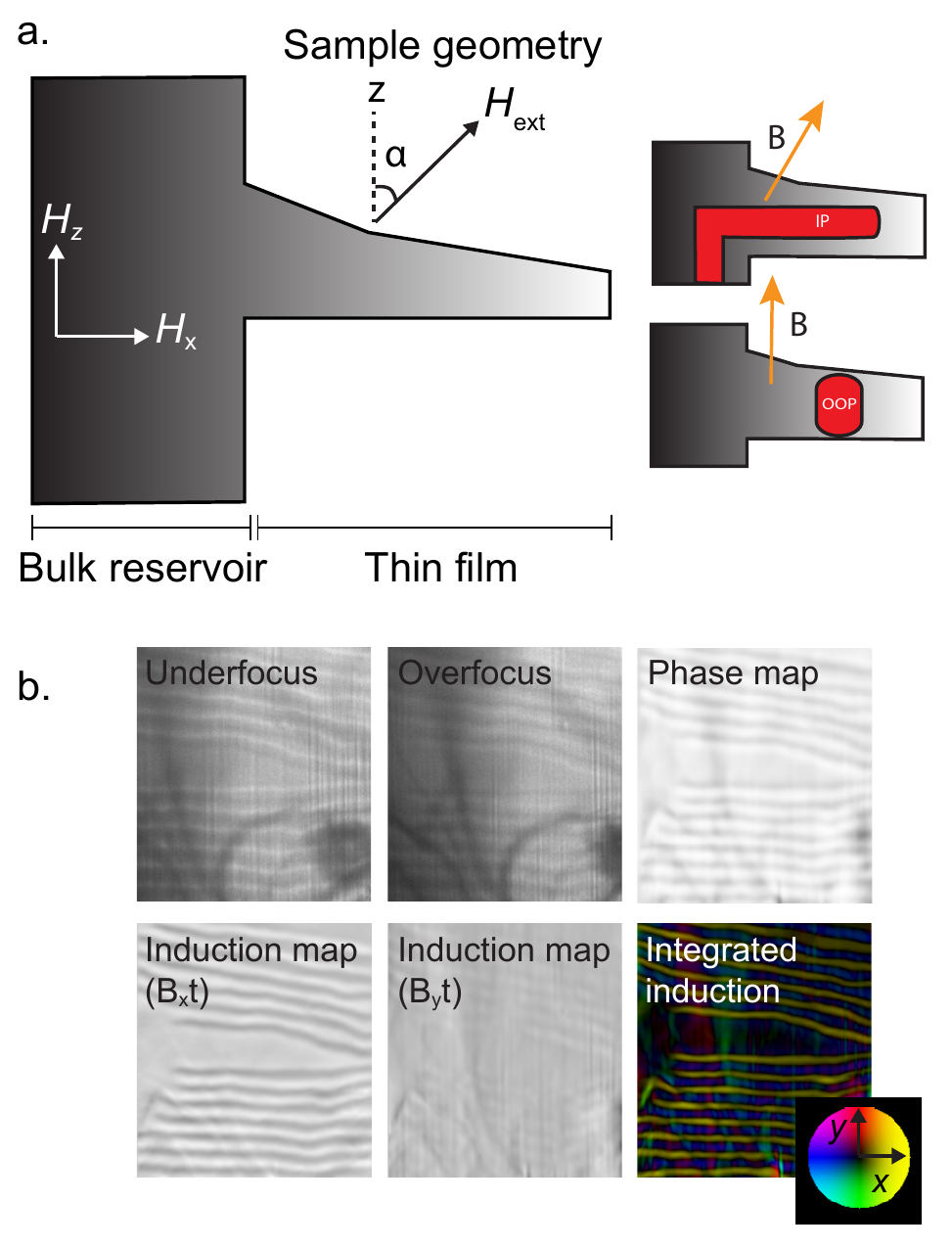}

\caption{\label{fig:2setup}(a) Schematic of the experimental geometry. The angle, $\alpha$, is defined as the angle between the film normal and the applied field. (b) Underfocus and overfocus LTEM image showing the reconstruction of magnetic phase.}
\end{figure}

To demonstrate the ability to image skyrmions in Co-Zn-Mn, we reconstruct the in-plane magnetization of the sample from a defocus series of LTEM images under constant conditions for the sample geometry shown in Figure \ref{fig:2setup}a. The underfocus and overfocus LTEM images in Figure \ref{fig:2setup}b were collected at $\alpha = 9\degree$ and an applied field strength of 120 mT in the direction shown in Figure \ref{fig:2setup}a. 
The horizontal textures in these images results from in-plane magnetic fields, which deflect the transmitted electron beam to create contrast. 
The phase map is created by subtracting the two defocus images to obtain the magnetization in the plane of the sample.
The integrated induction map generated using a phase retrieval algorithm includes contrast from the in-plane magnetic fields, where the inset color wheel indicates the local direction of the field. In this image, the horizontal contrast is consists of stripes of yellow separated by larger stripes of blue, with a periodicity of 140 nm in the y dimension. It can be inferred that these are IP skyrmions whose cores are oriented in the $-x$ direction, which opposes the direction of the in-plane applied magnetic field $H_x$ in the $+x$ direction. The skyrmion cores are separated by thin regimes with ferromagnetic character, where the in-plane field aligns with the applied field in the $+x$ direction. The magnetic phase reconstruction confirms the skyrmion-like character.

\FloatBarrier